\documentclass[aps,prd,shownopacs,groupedaddress,floatfix,twocolumn,nofootinbib,preprintnumbers]{revtex4}

\usepackage{graphicx}
\usepackage{bbm}
\usepackage{epsfig}
\usepackage{amsmath}
\usepackage{amssymb}
\usepackage{graphicx}
\usepackage{epstopdf}
\DeclareGraphicsExtensions{.eps, .eps, .jpg, .png}

\begin{document}

\preprint{ACFI-T14-24}


\title{Massless gravitons on a brane \\from massive gravity in five-dimensional Minkowski space}


\author{Basem Kamal El-Menoufi and Lorenzo Sorbo}

\affiliation{Department of Physics, University of Massachusetts\\ Amherst, MA 01003, USA}


\date{\today}

\begin{abstract}

We discuss the possibility of localizing Einstein gravity on a three-brane embedded in a five-dimensional Minkowski space by giving a space-dependent mass to the graviton. We show that, with an overall fine tuning of the mass profile, it is possible to localize a massless graviton with a finite mass gap from the massive Kaluza-Klein modes. The theory does not contain ghosts and no scalar massless mode is sourced by  brane matter. A massless vector mode is strongly coupled to the flux of energy-momentum into the extra dimension, so that it is not sourced as long as only brane-localized matter is considered. In the case where the mass profile has a delta-like shape the four-dimensional Planck mass $M_4$ is related to the five-dimensional one, $M_5$, by $M_4^{2}=M_5^3/m_0$, where $m_0$ is the mass of the first Kaluza-Klein mode of the graviton.

\end{abstract}


\maketitle


\section{Introduction}

There are well known field-theoretical mechanisms of localization of matter on a brane. One of the first brane models~\cite{Rubakov:1983bb} was based on a mechanism of localization of fermion zero modes on a domain wall generated by a field coupled to fermions via a Yukawa coupling~\cite{Jackiw:1975fn}. In that model fermions were getting an effective space dependent mass and a chiral zero mode was localized on the hypersurface where its  effective five-dimensional mass vanishes. 

Mechanisms analogous to that of~\cite{Rubakov:1983bb}, generalized to scalar fields and gauge bosons, were reviewed in~\cite{ArkaniHamed:1998rs}. In all these cases an effective space-dependent mass was leading to localization of the lowest-lying mode near the local minimum of the effective mass\footnote{In the case of gauge bosons the most natural way of generating such a mass is to give a space-dependent vacuum expectation value to a Higgs field. In this case, however, the four dimensional photon does not remain massless: the charge condensate off the brane screens charges on the brane, leading to a Yukawa-like four-dimensional interaction~\cite{ArkaniHamed:1998rs}. A way out is to consider a non abelian strongly coupled theory in the bulk that breaks down to an abelian massless gauge field on the brane~\cite{Dvali:1996xe}.}.

It is more difficult to localize gravity on a brane. The most widely studied model that is able to accomplish this is the Randall-Sundrum model~\cite{Randall:1999vf}, where a brane marks the fixed point of a $Z_2$ symmetry between two identical Anti-de Sitter bulk spaces. The bulk curvature accelerates away from the brane the  gravitons whose momentum has a component normal to it. As a consequence the wave function of these massive Kaluza-Klein modes has small overlap with the brane, effectively leaving only the zero mode in interaction with brane matter. Another model leading to four dimensional gravity with a noncompact (and, in this case, flat) bulk is the Dvali-Gabadadze-Porrati (DGP) model~\cite{Dvali:2000hr}, where the graviton has a space-dependent normalization of the kinetic term that leads to a metastable zero mode  on the brane.

In the present paper we discuss an alternative to the Randall-Sundrum and DGP models that is closer to the original ideas for localization discussed in~\cite{Rubakov:1983bb}. We will consider a five-dimensional theory of massive gravity where the graviton lives in a flat space-time and has a mass $m^2=m^2(w)$ that depends on the extra dimensional coordinate $w$ in a non trivial way, so that a zero mode is localized on our four-dimensional Universe.

Such a construction requires a fine-tuning, as the function $m^2(w)$ has to be designed in such a way that a zero mode is supported. It is worth noting, however, that if the bulk graviton mass comes from some superpotential $W(w)$ such that $m^2(w)\sim W'(w)+W(w)^2$, then the existence of a zero mode is guaranteed.

In order to lead to a zero mode, the mass profile of the graviton has to take negative values in some part of space. A constant negative mass of the graviton, however, is known to lead to a ghostlike degree of freedom (see, {\em e.g.},~\cite{ArkaniHamed:2002sp}). Therefore one might worry that our construction leads to a localized, ghostlike instability (see~\cite{Batell:2006dp} for a related discussion in the case of abelian gauge fields). We show that such a worry is not justified in our setting. In fact, a ghost-like degree of freedom with (imaginary) mass $m$ needs a region of size $\ell\gtrsim |m|^{-1}$ to develop its instability. The very requirement that the four-dimensional effective theory contains a zero mode and no tachyonic modes leads precisely to the constraint $\ell\ll |m|^{-1}$, and no ghost mode is allowed. 

A related issue is the possible presence of strongly coupled modes. Our construction displays such a mode, a vector  that however couples only to matter with a nonvanishing flow of energy-momentum towards the extra dimension. In the regime we are interested in, where matter stays localized on the brane, this mode is not sourced and is harmless.

As we will see, the scalar, vector and tensor components of the five-dimensional graviton obey different equations of motion. We will be able to find explicit solutions in the case where the mass profile has the shape of a square well. In particular, we will show explicitly that, for appropriate choices of the function $m^2(w)$, the four dimensional effective theory contains a zero mode for the graviton and a finite mass gap, showing that our model, for sufficiently large values of the scale that controls the bulk graviton mass, is compatible with four-dimensional massless gravity.

The construction of brane models is usually complicated by the existence of the dilaton, the four-dimensional scalar component of the five-dimensional graviton. If such a mode is massless and coupled with gravitational strength to the brane stress-energy tensor, it induces a deflection of light that is in disagreement with observations. Usually the dilaton is made massive in order to satisfy phenomenological constraints. In our system, as we will see, such a massless mode does not couple to conserved brane matter and has no observable effect. Therefore, in this regard, our model is phenomenologically safe.

To  our knowledge, this is the first work where the Kaluza-Klein decomposition of a massive graviton with a nontrivial, five-dimensional mass profile with finite thickness is analyzed. A similar analysis has been performed, in the limit of a $\delta$-like mass profile and in the context of a compact and curved extra dimension, in~\cite{Gherghetta:2005se}. The focus of our work is different from that of~\cite{Gherghetta:2005se}, as the main question we are asking concerns the possibility of localizing gravity in non compact five-dimensional Minkowski space. Our results agree with those of~\cite{Gherghetta:2005se}, where applicable. 

The plan of our work is the following. In section II we write down the linearized equations of motion of the five-dimensional massive graviton, we decompose the graviton into four-dimensional scalar, vector and tensor modes, and we write the equations of motion for those modes separately. In section III we write the four-dimensional effective action for the system, including the couplings of the Kaluza-Klein modes of tensors, vectors and scalars to matter. In section IV we derive explicitly the expression of the profiles of the five-dimensional wave functions in the case where the mass $m^2(w)$ takes the form of a square well. Section V deals with the special case of massive scalar modes in the square-well mass profile, that need a separate treatment. In section VI we consider the limit where the square-well mass profile is approximated by a Dirac $\delta$-function, and we compute the four-dimensional Planck scale as a function of the parameters of the five-dimensional theory. Section VII contains our conclusions and the discussion of future directions.

\section{perturbations around flat space and graviton decomposition}%

In order to study the possibility of localizing graviton fluctuations on a three-brane in a five-dimensional Minkowski bulk we will focus on the Pauli-Fierz Lagrangian where the mass $m^2(w)$ of the graviton depends on the extra coordinate $w$.  By expanding around flat space $g_{AB} = \eta_{AB} + h_{AB}$, our equations of motion are
\begin{align}
G^L_{AB} + m^2(w) M^L_{AB} = 0\,,
\end{align}
where $G^L_{AB}$ refers to the linearized Einstein tensor and $M^L_{AB}$ is the tensor resulting from varying the Pauli-Fierz mass term $\propto h_{AB}\,h^{AB}-h^2$
\begin{align}
M^L_{AB} = \frac{1}{2}(h_{AB} - \eta_{AB} h), \quad h = \eta^{AB} h_{AB}\,.
\end{align}
We next decompose the graviton field into modes that transform as tensors, vectors and scalars\footnote{In principle the zero modes should be treated separately as the decomposition~(\ref{metric pert}) becomes  in this case ambiguous, see for instance~\cite{Kiritsis:2006ua}. Nevertheless, we expect our analysis to provide the correct behavior of the system even in the case of the zero modes.} under $SO(3,\,1)$
\begin{align}\label{metric pert}
\nonumber
&h_{\alpha\beta} = \gamma_{\alpha\beta} + 2\, \partial_{(\alpha} E_{\beta)} + 2\,\psi\, \eta_{\alpha\beta} - 2\, \partial_{\alpha}\partial_{\beta} E\,, \\
&h_{w\alpha} = -\partial_{\alpha}B - B_\alpha\,, \quad h_{ww} = 2\, \phi \,,
\end{align}
where greek indices run over $0,1,2,3$ and where $\partial_\alpha \gamma^\alpha_{\beta} =\gamma^\alpha_{\alpha} =\partial^\alpha E_\alpha =\partial^\alpha B_\alpha=0$.  Employing this decomposition we obtain
\begin{align}
\nonumber
&G^L_{\alpha\beta}=-\frac{1}{2}\left(\Box\gamma_{\alpha\beta}+\gamma_{\alpha\beta}''\right)-\partial_{(\alpha} {\bf B}^\prime_{\beta)}, \\
&\phantom{G^L_{\alpha\beta}=}- \partial_\alpha \partial_\beta \left(2\,\psi+\Phi\right)+\eta_{\alpha\beta}\left[2\,\Box\psi+3\,\psi''+\Box\Phi \right]\nonumber\\
&G^L_{w\alpha}=\frac{1}{2}\left[\Box\,{\bf {B}}_\alpha-6\,\partial_\alpha\psi'\right], \quad G^L_{ww}=3\,\Box\psi\,,
\end{align}
where a prime denotes derivative with respect to the extra dimensional coordinate $w$, $\Box$ represents the four dimensional d'Alembert operator,  and where $\Phi = \phi + B^\prime - E^{\prime\prime}$ and ${\bf B}_\alpha\equiv B_\alpha-E'_\alpha$. Moreover
\begin{align}
\nonumber
&M^L_{\alpha\beta}=\frac{1}{2}\gamma_{\alpha\beta}+\partial_{(\alpha}E_{\beta)} - \partial_\alpha \partial_\beta E +\eta_{\alpha\beta}\left(-3\,\psi+\Box E-\phi\right),\\
&M^L_{w\alpha}=-\frac{1}{2}\left(\partial_\alpha B+B_\alpha\right), \quad M^L_{ww}=-4\,\psi+\Box E\,.
\end{align}
Note that the mass term breaks invariance under diffeomorphisms. In its absence, gauge invariant variables would be given by $\gamma_{\alpha\beta}$, ${\bf B}_\alpha$, $\psi$ and $\Phi$. 

Due to the transformation properties of each mode, the equations of motion decouple into three sectors: tensor, vector and scalar.

\subsection{Tensors}

The equation for the tensor mode, which is to be interpreted as the four-dimensional graviton, reads 
\begin{align}
\Box\gamma_{\alpha\beta}+\gamma_{\alpha\beta}''-m^2(w)\,\gamma_{\alpha\beta} = 0\,.
\end{align}

The equation is separable, so that we write
\begin{equation}
\gamma_{\alpha\beta}(x^\lambda,\,w)=\sum_n\bar\gamma_{\alpha\beta}^{(n)}(x^\lambda)\,f^{(n)}_\gamma(w)\,,
\end{equation}
where
\begin{align}
&f^{(n)}_\gamma{}''(w)-m^2(w)\,f^{(n)}_\gamma(w) = -\left(\mu_n^\gamma\right)^2\,f^{(n)}_\gamma(w)\,,\nonumber\\
&\Box\bar\gamma_{\alpha\beta}^{(n)}(x^\lambda)=\left(\mu_n^\gamma\right)^2\,\bar\gamma_{\alpha\beta}^{(n)}(x^\lambda)\,,
\end{align}
with $\left(\mu_n^\gamma\right)^2$ a separation constant.

\subsection{Vectors}

There are two coupled equations for the vector modes
\begin{align}\label{eqvec}
&B'_{\alpha} + E_\alpha''-m^2\,E_\alpha =0\,\nonumber\\
&(\Box - m^2) B_\alpha + \Box E^\prime_\alpha  =0\,.
\end{align}

By introducing the combination ${\bf{B}}_\alpha \equiv B_\alpha + E'_\alpha$ ({\em i.e.}, the degree of freedom that would be gauge invariant in the massless theory), we can write the equations of motion as
\begin{align}\label{mastervec}
{\bf{B}}''_{\alpha} - \frac{(m^2)'}{m^2}{\bf{B}}'_\alpha + (\Box-m^2) {\bf{B}}_\alpha = 0, \quad E_\alpha =\frac{{\bf{B}}'_\alpha}{m^2}\,.
\end{align} 
The equation for ${\bf B}_\alpha$ is separable, so we can write its solutions as
\begin{equation}
{\bf{B}}_\alpha(x^\lambda,\,w)=\sum_n \bar{B}_\alpha{}^{(n)}(x^\lambda)\,f^{(n)}_B(w)\,,
\end{equation}
where $f^{(n)}_B$ satisfies the ordinary differential equation
\begin{align}
f^{(n)}_B{}^{\prime\prime} - \frac{(m^2)'}{m^2} \,f^{(n)}_B{}^{\prime}- m^2\,f^{(n)}_B = -\left(\mu_n^B\right)^2\,f^{(n)}_B\,.
\end{align}
 
Note that the vector and the tensor modes satisfy different equations of motion, and, as a consequence, they generally have different mass spectra $\mu_n^B\neq\mu_n^\gamma$.
 
\subsection{Scalars}

There are four equations for the scalar modes
\begin{align}
&2\,\psi + \Phi + m^2(w)\,E=0\,,\nonumber\\
&2\,\Box\psi + 3\,\psi'' + \Box \, \Phi + m^2(w)\left(-3\,\psi + \Box E - \phi\right)=0\,,\nonumber\\
&-6\,\psi'-m^2(w)\,B=0\,,\nonumber\\
&3\,\Box\psi+m^2(w)\,\left(-4\,\psi+\Box E\right)=0\,,
\end{align}
that can be reduced to a single equation if we introduce the scalar degree of freedom $s$ that is related to $\psi$, $\phi$, $E$ and $B$ via
\begin{align}
&\psi=\frac{\Box s}{4}\,,\qquad E=s-\frac{3}{4\,m^2}\,\Box s\,,\qquad\\
&B=-\frac{3}{2\,m^2}\,\Box\,s'\,,\qquad \phi=E''-m^2\,E-B'-2\,\psi\,.\nonumber
\end{align}
The field $s$ satisfies the equation 
\begin{align}\label{eoms}
\left(1- \frac{3}{4}\left(m^{-2}\right)''\right)\,\Box s+s''- m^2\, s =0\,,
\end{align}
that turns out to be separable. To see this, let us write $s = W(w)\, X(x^\lambda)$, hence 
\begin{align}
X\, W^{\prime\prime} + W\, \Box X - m^2(w)\, W\, X - \left(\frac{3}{4m^2}\right)^{\prime\prime} W\, \Box X = 0
\end{align}
We divide by $s$ and $1-\left(3/4m^2\right)^{\prime\prime}$ to find
\begin{align}
\left[1-\left(\frac{3}{4m^2}\right)^{\prime\prime}\right]^{-1} \left(\frac{W^{\prime\prime}}{W} - m^2(w) \right) = - \frac{\Box X}{X} \,,
\end{align}
so that eq.~(\ref{eoms}) is separable, which allows us to decompose 
\begin{equation}
s(x^\lambda,\,w)=\sum_n \bar{s}^{(n)}(x^\lambda)\,f^{(n)}_s(w)\,,
\end{equation}
where $f^{(n)}_s$ satisfies the ordinary differential equation
\begin{align}\label{eqfs}
f_s^{(n)}{}''- m^2\, f_s^{(n)} =-\left(1- \frac{3}{4}\left(\frac{1}{m^{2}}\right)''\right)\,(\mu_n^s)^2\, f_s^{(n)}\,.
\end{align}

\section{The 4$D$ effective action}%

In this section we will first compute the free four-dimensional action for the Kaluza-Klein modes of the graviton, and then we will discuss the couplings of the various modes to matter.

\subsection{The free action}

The quadratic action reads
\begin{align}
&S^{\rm {free}}_{5D} =\frac{M_5^3}{2}\int d^4x\,dw \bigg[\frac{1}{4} h_{AB} \Box_5 h^{AB} + \frac{1}{2} h \, \partial_A \partial_B h^{AB} \nonumber\\
&- \frac{1}{4}h \, \Box_5 h - \frac{1}{2} h_{AB} \, \partial_{C} \partial^{B} h^{AC}-\frac{m^2}{4}\left(h_{AB}\,h^{AB}-h^2\right)\bigg]\,.
\end{align}
The actions for different modes decouple as expected. Integrating over $w$ yields the 4$D$ effective action for tensors
\begin{align}
S^{(2)}_{\rm {free}} = \frac{M_5^3}{8}\,\sum_n \int d^4x\, \bar\gamma^{(n)}{}^{\alpha\beta} \left( \Box - (\mu_n^\gamma)^2 \right) \bar\gamma^{(n)}_{\alpha\beta}\,.
\end{align}

In the case of vectors, after using the equations of motion the action reads
\begin{align}\label{effactvec}
&S^{(1)}_{\rm {free}} = \frac{M_5^3}{4}\,\sum_n\,(\mu_n^B)^2 \int d^4x\,  \bar{B}^{(n)}_{\alpha} \left(\Box - (\mu_n^B)^2 \right) \bar{B}^{(n)}{}^{\alpha},
\end{align}
where we have used the fact that the eigenfunctions $f^{(n)}_B$ are orthogonal with respect to the measure $m(w)^{-2}$ and we have normalized them according to:
\begin{equation}
\int \frac{dw}{m^2(w)}\,f^{(n)}_B(w)\,f^{(m)}_B(w)=\delta_{mn}\,.
\end{equation}
Note in particular the coefficient $(\mu_n^B)^2$ in front of the action of the $n$-th vector mode, that suggests that tachyon modes are actually ghostlike in our system. This is the four-dimensional counterpart of the fact that a tachyonic massive graviton is ghostlike in the full five-dimensional theory (see for instance~\cite{ArkaniHamed:2002sp}). The action~(\ref{effactvec}) shows that, even if the mass profile $m^2(w)$ is negative in some region of space, the theory does not contain ghostlike instabilities as long as there are no tachyon modes in the four-dimensional theory. It also shows that massless modes have a vanishing kinetic term, raising the worry that they might couple strongly to matter. We will discuss this issue in section III B., where we will see that as long as there is no flow of energy-momentum into the extra dimension the vector modes are not sourced.

The full action for the scalars reads
\begin{align}
&S^{(0)}_{\rm {free}} = \frac{M_5^3}{2}\int d^4x\,dw \bigg[ - 6\, \psi\,\left(\Box \psi+\Box \,\Phi + 2\, \psi''\right) \\
& \left.+m^2\left(12\,\psi^2+\frac{1}{2} B\,\Box B - 6\, \psi\, \Box E + 8\, \psi\, \phi - 2\,\phi\, \Box E\right)\right]\,. \nonumber
\end{align}
By using the scalar equation of motion it is possible to write the free action solely in terms of $\psi$
\begin{align}
S^{(0)}_{\rm {free}} =12 \int d^4x\,dw &\Big[\psi\,\Box\psi-\psi'{}^2\nonumber\\
&\left. -\psi\left(m^2+\frac{3}{4} \left(\frac{\Box}{m^2}\right)''\right) \psi\right]\,.
\end{align}
Trading $\psi$ for $s$, and introducing the Kaluza-Klein decomposition for $s$, the action can be written as
\begin{eqnarray}\label{action4s}
S^{(0)}_{\rm {free}}&=\frac{3}{4}\sum_{n,n'}\int dw\,\left(1-\frac{3}{4}\left(\frac{1}{m^2}\right)''\right)\,f^{(n)}_s(w)\,f^{(n')}_s(w)\nonumber\\
&\times(\mu_n^s)^2\,(\mu_{n'}^s)^2\,\int d^4x\, \bar{s}^{(n)} \left( \Box - \mu_n^2 \right) \bar{s}^{(n')}\,.
\end{eqnarray}
and it is possible to show that the functions $f^{(n)}_s(w)$ are orthogonal with respect to the measure $\left(1-\frac{3}{4}\left(\frac{1}{m^2}\right)''\right)$. Therefore we can set\footnote{Since the function$\left(1-\frac{3}{4}\left({m^{-2}}\right)''\right)$ is not positive everywhere, one should check that the integral appearing in eq.~(\ref{norms}) is actually positive if we want to avoid ghost modes. This turns out to be the case for a specific mass profile $m^2(w)$ that we will consider below.} 
\begin{eqnarray}\label{norms}
\int dw\,\left(1-\frac{3}{4}\left(\frac{1}{m^2}\right)''\right)\,f^{(n)}_s(w)\,f^{(n')}_s(w)=\frac{2}{3}\delta_{nn'}\,,
\end{eqnarray}
so that
\begin{align}\label{action4s1}
S^{(0)}_{free}=\frac{1}{2}\sum_{n}(\mu_{n}^s)^4\,\int d^4x\, \bar{s}^{(n)} \left( \Box - \mu_n^2 \right) \bar{s}^{(n)}\,.
\end{align}
Note that we must make sure that the function $f^{(n)}_s(w)$ is square integrable with respect to the measure $\left(1-\frac{3}{4}\left({m^{-2}}\right)''\right)$. As we will see in section V, this requirement imposes important restrictions on the spectrum of scalar perturbations in the theory.

\subsection{Coupling to matter}

The part of action describing the interaction with matter reads
\begin{align}
S_{5D}^{\rm {int}} = \int d^4x\, dw\, h_{AB}\, T^{AB}\,,
\end{align}
where $T_{AB}$ is the energy-momentum tensor which we assume to be factorizable
\begin{align}
T_{AB} = T^0_{AB} \mathcal{T}(w)\,.
\end{align}

In case of matter localized on a singular brane, $\mathcal{T}(w)=\delta(w)$  and $T_{\mu w}=T_{ww}=0$. We now decompose the energy-momentum tensor as follows
\begin{align}
T^0_{\mu\nu} &= t_{\mu\nu} + 2\, \partial_{(\mu} t^E_{\nu )} + 2\, T^s\, \eta_{\mu\nu} - 2\, \partial_\mu \partial_\nu t^E\,, \\
T^0_{\mu w} &= -\partial_\mu t^B - t^B_\mu \, , \quad T^0_{ww} =  2 \,p\,,
\end{align}
where $t_\mu ^{~\mu} = \partial_\mu t^{\mu\nu} = \partial^\mu t^E_\mu = \partial^\mu t^B_\mu = 0$. 

Stress-energy conservations leads to the following identities:
\begin{align}\label{emcons}
\nonumber
& 2\,T^s - 2\, \Box t^E - t^{B \prime} = 0\, \\
\nonumber
&\Box t^E_\mu - t_\mu ^{B \prime} = 0\,, \\
&\Box t^B -2\, p^\prime =0 \,,
\end{align}
so that in the limit of a $\delta$-like brane, $p=t^B=t_\mu^B=0$, one has $\Box t^E_\mu=0$ and $T^s=\Box t^E$.

Integrating over the extra dimension, we find the 4$D$ interaction terms
\begin{align}
&S^{(2)}_{\rm {int}} =  \sum_n g^{(2)}_n \int d^4x \,  t^{\mu\nu}\, \bar\gamma^{(n)}_{\mu\nu}\,,\\
&S^{(1)}_{\rm {int}} =  2\, \sum_n g^{(1)}_n \int d^4x \, t^B{}^\mu\,{\bar {B}}^{(n)}_\mu \,,
\end{align}
where we used the conservation of the energy-momentum tensor and where we have defined
\begin{align}
g^{(2)}_n &= \int dw \, f_\gamma^{(n)}(w)\, \mathcal{T}(w) \\
g^{(1)}_n &= \int dw\,f_B^{(n)}(w)\, \mathcal{T}(w)
\end{align}
The scalar interaction term reads
\begin{align}
\nonumber
S^{(0)}_{\rm {int}} = 4\,\int d^4x\, dw \, \left[\left(4\,T^s - \Box t^E\right) \psi+p\,\Phi\right]\,,
\end{align}
where once again we used Eq.~(\ref{emcons}). 

In particular, in the limit of a $\delta$-like brane the coupling to matter reduces to
\begin{align}
\nonumber
S_{\rm {int}}^{(0)} =\sum_n\int d^4x &\left(f_\gamma^{(n)}(0)\,t_{\mu\nu}\, \bar\gamma_{(n)}^{\mu\nu}\right.\nonumber\\
&+\left.3\,(\mu_n^s)^2\,f_s^{(n)}(0)\,T^s\,\bar{s}^{(n)}\right)\,,
\end{align}
that, together with eq.~(\ref{action4s}), gives the important result that the zero mode of the scalars decouples completely from the action, making the theory phenomenologically safe for what concerns the constraints from light bending. Moreover, also the zero mode of the vector $\bar{B}_\alpha$ decouples in this regime, so that, at least at the linearized level and in the absence of flow of matter along the fifth dimension, there are no concerns of strong coupling even if the kinetic term of the zero mode of $\bar{B}_\alpha$ is vanishing.

\section{An exactly solvable mass profile: the square well}%

In this section, we solve the Schr\"odinger-like equations for a thick brane with a square-well shape of width $2\,w_0$:
\begin{align}\label{square}
m^2(w) = \left\{
\begin{array}{ll}
m_0^2\,, & |w|>w_0\\
 - m_1^2\,, & |w|<w_0\end{array}
 \right.\,,
\end{align}
where we will determine the relation between $m_0$, $m_1$ and $w_0$ by demanding the existence of a zero mode and no tachyons.

\subsection{Tensors}

The equation for the tensor mode is the standard Schr\"odinger equation for a particle in a square well. For the case at hand, bound states exist as long as $-m_1^2 < \mu^2 < m_0^2$. The even-parity solution reads
\begin{align}\label{gammasol}
f_\gamma(w) \propto
\left\{ \begin{array}{ll}
\cos\left(k\, w\right), &\quad |w|<w_0 \\
e^{-\kappa \,|w|}, &\quad |w|>w_0
\end{array}
\right.\,,
\end{align}
where
\begin{align}\label{defk}
k \equiv \sqrt{m_1^2+\mu^2}, \quad \kappa \equiv \sqrt{m_0^2 - \mu^2}\,.
\end{align}
Matching the solution and its first derivative at $w_0$ yields a transcendental equation for the eigenvalues
\begin{align}
\text{tan} (z) = \sqrt{\left(\frac{z_0}{z}\right)^2-1}
\end{align}
where
\begin{align}
z=k\, w_0, \quad z_0 = \sqrt{m_0^2 + m_1^2}\,w_0\,.
\end{align}

The condition for the existence of a zero mode imposes 
\begin{align}\label{zm}
m_0 = m_1 \, \tan\left(m_1\,w_0\right)\,,
\end{align}
and the condition that there is only one bound, massless, parity-even state\footnote{We will assume that the matter distribution along the extra dimension is even under $w\leftrightarrow -w$, so that graviton modes that are odd under $w\leftrightarrow -w$ do not couple to matter and can be disregarded.} (that implies, in particular, the absence of tachyons) yields the inequality
\begin{equation}\label{condonlyzero}
m_1\,w_0<\pi\,\cos\left(m_1\,w_0\right)\,.
\end{equation}

On the top of the isolated zero mode discussed above, the four-dimensional effective theory then displays a continuum of massive modes with $\mu^2>m_0^2$. Focusing on the even-parity solutions, we get
\begin{align}\label{gammamassive}
f_\gamma(w) \propto
\left\{ \begin{array}{ll}
\cos\left(\sqrt{\mu^2 + m_1^2}\, w \right), &\quad |w|<w_0 \\
\cos\left(\sqrt{\mu^2 - m_0^2}\, |w| + \alpha\right), &\quad |w|>w_0
\end{array}
\right.\,,
\end{align}
where $\alpha$ is a phase to be determined from the matching conditions
\begin{align}
\kappa^\prime \,\tan(\kappa^\prime\, w_0 + \alpha) = k\, \tan(k\,w_0)\,,
\end{align}
where $\kappa^\prime \equiv\sqrt{\mu^2-m_0^2}$.

\subsection{Vectors}

In the case of vectors, it is convenient to solve the original system of equations~(\ref{eqvec}). In terms of the variable ${\bf{B}}_\alpha\equiv B_\alpha+E_\alpha'$, after separating variables as ${\bf{B}}_\alpha(w,\,x^\lambda)=f_B(w)\,\bar{B}_\alpha(x^\lambda)$ and $E_\alpha(w,\,x^\lambda)=f_E(w)\,\bar{E}_\alpha(x^\lambda)$ with $\Box \,\bar{B}_\alpha(x^\lambda)=\mu^2\,\bar{B}_\alpha(x^\lambda)$ and $\Box \,E_\alpha(x^\lambda)=\mu^2\,E_\alpha(x^\lambda)$ the relevant equations read
\begin{align}\label{eqvecred}
&f_B'(w)-m^2(w)\,f_E(w)=0\nonumber\\
&\mu^2\,f_B(w)-m^2(w)\,\left(f_B(w)-f_E'(w)\right)=0\,,
\end{align}
that are solved for $\mu^2<m_0^2$ by
\begin{align}
&f_B(w)=\left\{
\begin{array}{ll}
c_o\,{\mathrm {sign}}(w)\,e^{-\kappa\,|w|} & |w|>w_0 \\
c_i\,\sin\left(k\,w\right) & |w|<w_0
\end{array}
\right.\,,\nonumber\\
&f_E(w)=\left\{
\begin{array}{ll}
-\frac{\kappa}{m_0^2}\,c_o\,e^{-\kappa\,|w|} & |w|>w_0 \\
-\frac{k}{m_1^2}\,c_i\,\cos\left(k\,w\right) & |w|<w_0
\end{array}\right.\,,
\end{align}
where $k$ and $\kappa$ are defined in eq.~(\ref{defk}) and where $c_o$ and $c_i$ are integration constants. Notice that we focus only on the situation where $f_B(w)$ is odd and $f_E(w)$ is even under $w\leftrightarrow -w$, since the in the opposite case the graviton modes do not couple to brane matter that even under $w\leftrightarrow -w$.
Inspection of eqs.~(\ref{eqvecred}) shows that both $f_B(w)$ and $f_E(w)$ must be continuous when $w$ crosses $w_0$, implying the following condition on the parameter $\mu$:
\begin{align}
\tan\left(z\right)=\frac{m_0^2/m_1^2}{\sqrt{\frac{z_0^2}{z^2}-1}}\,,
\end{align}
that in general leads to a mass spectrum that is different from that of tensors. Once the condition~(\ref{zm}) for the existence of a tensor zero mode $\mu=0$ is imposed, also the vectors will have a zero mode. For $m_1\,w_0$ sufficiently smaller than unity (that corresponds to the regime where also condition~(\ref{condonlyzero}) is satisfied), one can see that actually {\em only} the zero mode will be allowed in the entire range $-m^2_1<\mu^2<m^2_0$. In particular, this implies that no tachyons (and therefore no ghosts, as we have discussed in section III A. above) will be allowed.

For $\mu^2>m_0^2$ the solutions to eqs.~(\ref{eqvecred}), with $f_E$ even with respect to $w\to-w$ are given by
\begin{align}
&f_B(w)=\left\{
\begin{array}{ll}
d_o\,{\rm {sign}}(w)\,\sin\left(\kappa'\,|w|+\alpha\right)& |w|>w_0 \\
d_i\,\sin\left(k\,w\right) & |w|<w_0
\end{array}
\right.\,,\nonumber\\
&f_E(w)=\left\{
\begin{array}{ll}
\frac{\kappa'}{m_0^2}\,d_o\,\cos\left(\kappa'\,|w|+\alpha\right) & |w|>w_0 \\
-\frac{k}{m_1^2}\,d_i\,\cos\left(k\,w\right) & |w|<w_0
\end{array}\right.\,,
\end{align}
where $d_i$, $d_o$ and $\alpha$ are integration constants.

By imposing continuity of $f_E$ and of $f_B$ at $|w|=w_0$ we obtain the following expression for $\alpha$
\begin{align}
\tan\left(\kappa'\,w_0+\alpha\right)=-\frac{\kappa'\,m_1^2}{k\,m_0^2}\,\tan\left(k\,w_0\right)\,.
\end{align}

\subsection{Scalars}

In the case of scalars the derivation of the profile of the zero mode is straightforward and we will present it here. The analysis of the massive modes is more elaborate, and we will defer it to the next section. 

Once we set $\mu_{n=0}^s=0$, eq.~(\ref{eqfs}) reduces to
\begin{align}\label{eoms0}
f_s^{(0)}{}''- m^2\, f_s^{(0)} =0\,,
\end{align}
which is the same as the equation for the zero mode of the tensor, so it has the same solution~(\ref{gammasol}). In terms of the variables appearing in the metric perturbation, this leads to
\begin{equation}
\psi=\phi=B=0\,,\,\qquad E=f^\gamma_{\mu=0}(w)\,\bar{E}\left(x^\lambda\right)\,,
\end{equation}
where $f^\gamma_{\mu=0}(w)$ is obtained by setting $\mu=0$ in eq.~(\ref{gammasol}) and where $\bar{E}\left(x^\lambda\right)$ satisfies $\Box\bar{E}\left(x^\lambda\right)=0$.

As one can see from the analysis of section III, the zero mode $E$, even if it is non vanishing, does not contribute to the four dimensional effective action. As a consequence our model does not display any physical massless scalar degree of freedom.

\section{Massive scalar modes}

In order to discuss the existence of massive scalar modes we need to consider a regularized version of the mass profile $m^2(w)$. In fact the function $(m^{-2}(w))''$ is ill-defined if $m^2(w)$ contains step functions as it does in eq.~(\ref{square}). We will therefore consider the following regularized mass profile for the scalar modes
\begin{align}\label{contm2}
m^2(w)=\left\{
\begin{array}{ll}
-m_1^2\,, & |w|<w_0-\frac{m_1^2}{\Lambda^3}\\
\Lambda^3\left(|w|-w_0\right)\,, & w_0-\frac{m_1^2}{\Lambda^3}<|w|<w_0+\frac{m_0^2}{\Lambda^3}\\
m_0^2\,, & |w|>w_0+\frac{m_0^2}{\Lambda^3}
\end{array}
\right.\,,
\end{align}
where we will be interested in the limit of $\Lambda\to\infty$, where we recover the square well profile~(\ref{square}). 

At the point $w=w_0$, $(m^{-2})''$ diverges as $(w-w_0)^{-3}$, corresponding to an essential singularity in the equation for $f_s$. As a consequence we treat the problems with $|w|>w_0$ and $|w|<w_0$ separately. 

Let us start with $|w|>w_0$ and let us choose $w>w_0$ (an analogous discussion will apply to the case $w<-w_0$). In this region we have
\begin{align}\label{1om2sec}
\left(\frac{1}{m^2}\right)''=2\,\frac{\Theta(w_0+\frac{m_0^2}{\Lambda^3}-w)}{\Lambda^3\,\left(w-w_0\right)^3}+\frac{\Lambda^3}{m_0^4}\delta(w_0+\frac{m_0^2}{\Lambda^3}-w)\,.
\end{align}

Let us first focus on the regime $\mu^2<m_0^2$. In the region $w>w_0+\frac{m_0^2}{\Lambda^3}$, the normalizable solution of eq.~(\ref{eqfs}) is 
\begin{equation}
f_s\left(w>w_0+\frac{m_0^2}{\Lambda^3}\right)\propto e^{-\sqrt{m_0^2-\mu^2}\,|w|}\,.
\end{equation}
In the region with $w_0<|w|<w_0+\frac{m_0^2}{\Lambda^3}$ the Schr\"odinger equation, after defining a new variable $\tilde{w}$ as $w=w_0+\frac{\mu^2}{\Lambda^3}\tilde{w}$, reads
\begin{equation}\label{eqswtilde}
\frac{d^2f_s}{d\tilde{w}^2}+\left[\frac{\mu^6}{\Lambda^6}\left(1-\tilde{w}\right)-\frac{3}{2\,\tilde{w}^3}\right]\,f_s=0\,,
\end{equation}
with $0<\tilde{w}<\frac{m_0^2}{\mu^2}(>1)$ for $\mu^2>0$, and $\frac{m_0^2}{\mu^2}(<-1)<\tilde{w}<0$ for $\mu^2<0$. In the limit $\Lambda\to \infty$ the term in $\mu^6/\Lambda^6$ is negligible. For  $\mu^2>0$, the normalizable solution to the resulting equation $\frac{d^2f_s}{d\tilde{w}^2}-\frac{3}{2\,\tilde{w}^3}\,f_s=0$ is
\begin{equation}
f_s(0<\tilde{w}<\frac{m_0^2}{\mu^2})\propto\sqrt{\tilde{w}}\,K_1\left(\sqrt{6/\tilde{w}}\right)\,,
\end{equation}
where $K_1$ is the Bessel function of imaginary argument.

Finally, we need to match the $\delta$-like part of the Schr\"odinger equation that comes from the last term of eq.~(\ref{1om2sec}). This is obtained by imposing
\begin{align}
\lim_{\epsilon\to 0^+}&\left[f_s'\left(\left(w_0+\frac{m_0^2}{\Lambda^3}\right)+\epsilon\right)-f_s'\left(\left(w_0+\frac{m_0^2}{\Lambda^3}\right)-\epsilon\right)\right]\nonumber\\
&=\frac{3}{4}\,\mu^2\,\frac{\Lambda^3}{m_0^4}\,f_s\left(w_0+\frac{m_0^2}{\Lambda^3}\right),
\end{align}
that gives, after requiring continuity of $f_s$ at $w=w_0+\frac{m_0^2}{\Lambda^3}$, 
\begin{equation}
-\sqrt{m_0^2-\mu^2}-\sqrt{\frac{3}{2}}\,\frac{\Lambda^3\,\mu}{m_0^3}\,\frac{K_2(\sqrt{6}\,\mu/m_0)}{K_1(\sqrt{6}\,\mu/m_0)}=\frac{3}{4}\,\frac{\mu^2\,\Lambda^3}{m_0^4},
\end{equation}
that for $\Lambda\to\infty$ reduces to the equation
\begin{equation}
\frac{K_2(\Xi)}{K_1(\Xi)}=-\frac{\Xi}{4}\,,
\end{equation}
with $\Xi=\sqrt{6\,\mu^2/m_0^2}$, that has no root in the range $0<\mu^2<m_0^2$, as one can check numerically.

In the case $\mu^2<0$ the variable $\tilde{w}$ takes negative values, and the solution to eq.~(\ref{eqswtilde}) reads
\begin{equation}
f_s(\tilde{w})=c_J\,\sqrt{-\tilde{w}}\,J_1\left(\sqrt{-6/\tilde{w}}\right)+c_Y\,\sqrt{-\tilde{w}}\,Y_1\left(\sqrt{-6/\tilde{w}}\right)
\end{equation}
(where $J_1$ and $Y_1$ are Bessel functions), that for $\tilde{w}\to 0^-$ goes as 
\begin{align}
f_s(\tilde{w})\propto \tilde{w}&\left[\left(c_Y+c_J\right)\,\cos\left(\sqrt{-6/\tilde{w}}\right)\right.\nonumber\\
+&\left.\left(c_Y-c_J\right)\,\sin\left(\sqrt{-6/\tilde{w}}\right)\right]\,.
\end{align}

Since the measure according to which $f_s$ should be square-normalizable is $\left(1-\frac{3}{4}\left(m^{-2}\right)''\right)$, that goes as $|\tilde{w}|^{-3}$ as $\tilde{w}\to 0^-$, the eigenfunction in this case is never normalizable. As a consequence, the mode equation has no normalizable solution for $\mu^2<0$ in the region $|w|>w_0$. Together with the discussion of the case $\mu^2>0$ above, this leads to the conclusion that there are no acceptable solutions for $\mu^2<m_0^2$ to eq.~(\ref{eqfs}) in the region $|w|>w_0$

A very similar analysis can be performed in the region $|w|<w_0$, focusing only on the solutions that are even under $w\to -w$ (that are those that will couple to parity-even brane matter). There are two main differences. First, $\mu^2>0$ corresponds in this case to $\tilde{w}<0$. As a consequence, by applying the argument used in the region $|w|>w_0$, $\mu^2<0$, we see immediately that there are no normalizable solutions for any $\mu^2>0$ in the region $|w|<w_0$. For what concerns solutions with $-m_1^2<\mu^2<0$, a discussion analogous to the one of the case $\mu^2>0$, $|w|>w_0$ above leads to the result that the eigenvalue $\mu^2$ must satisfy the equation 
\begin{equation}
\frac{K_2(\Xi)}{K_1(\Xi)}=\frac{\Xi}{4}\,,\qquad \Xi\equiv\sqrt{-6\,\mu^2/m_1^2}
\end{equation}
that has no solution in the  region $-1<\mu^2/m_1^2<0$.

We therefore conclude that eq.~(\ref{eqfs}) does not admit any normalizable solution for $-m_1^2<\mu^2<m_0^2$.

On the other hand, there exist normalizable solutions for all values of $\mu^2>m_0^2$. By virtue of the above analysis, such solutions are confined to the region $|w|>w_0$ and we find
\begin{align}\label{scalarmassive}
f_s(w) \propto
\left\{ \begin{array}{ll}
 \sqrt{|\tilde{w}|} K_1(\sqrt{6/|\tilde{w}|}), &\quad w_0 < |w| < w_0 +\frac{m_0^2}{\Lambda^3}  \\
 \cos \left( \kappa^\prime |w| + \alpha \right) , &\quad |w|> w_0 + \frac{m_0^2}{\Lambda^3} \\
0, &\quad |w|<w_0
\end{array}
\right.
\end{align}
where all the variables are defined as before and $\alpha$ is an arbitrary phase to be determined by the matching conditions which yield
\begin{align}
-\kappa^\prime \tan\left[\kappa^\prime \left(w_0 + \frac{m_0^2}{\Lambda^3}\right) + \alpha \right] = \sqrt{\frac{3}{2}} \frac{\mu \Lambda^3}{m_0^3} \frac{K_2(\Xi)}{K_1(\Xi)} + \frac{3\mu^2 \Lambda^3}{4m_0^4} \,,
\end{align}
that in the limit $\Lambda \rightarrow \infty$ gives $\alpha=\pi/2-\kappa'\,w_0$, so that the solution, for $\mu^2>m_0^2$, reads simply
\begin{align}
f_s(w) \propto
\left\{ \begin{array}{ll}
\sin \left( \kappa^\prime (|w|-w_0) \right) , &\quad |w|> w_0 \\
0, &\quad |w|<w_0
\end{array}
\right. .
\end{align}
We note here that in~\cite{Gherghetta:2005se} it was concluded that no massive excitations of the scalar mode are allowed. Even if the situation is different in our case, where massive modes are allowed, our result does not contradict that of~\cite{Gherghetta:2005se}, since the system considered in that work was compact, and therefore one had to satisfy boundary conditions at the two boundaries of the Randall-Sundrum brane, whereas in our case we are considering mode functions that have a fixed boundary condition only at the location of the brane, while at infinity we have plane waves.

\section{A special case:  $\delta$-like mass profile and the four-dimensional Planck mass}%

In this section we consider the limit of a $\delta$-like mass profile
\begin{equation}
m^2(w)=m_0^2-\tilde{m}_1\,\delta(w)\,,
\end{equation}
whose solutions can be obtained from the square well results discussed previously by taking the limit $w_0\to 0$, $m_1\to\infty$ with $\tilde{m}_1=2\,w_0\,m_1^2$ finite. The condition~(\ref{zm}) for the existence of a zero mode gives in this limit $\tilde{m}_1=2\,m_0$.

Let us now review the expressions of the (normalized) mode functions in this regime.

\subsubsection{Tensors}

The zero mode is given by
\begin{equation}
f_\gamma^0(w)=\sqrt{m_0}\,e^{-m_0\,|w|}\,,
\end{equation}
whereas the massive modes are given by
\begin{align}
f_\gamma^\mu(w)=&\sqrt{\frac{\,\mu}{\pi\,\sqrt{\mu^2-m_0^2}}}\,\cos\left(\sqrt{\mu^2-m_0^2}|w|+\alpha\right)\,,\nonumber\\
\cos\alpha=&\frac{\sqrt{\mu^2-m_0^2}}{\mu}\,,
\end{align}
and are normalized according to $\int dw\,f^{\mu}_\gamma(w)\,f^{\mu'}_\gamma(w)=\delta(\mu-\mu')$.

\subsubsection{Vectors}

The canonically normalized zero mode is given by
\begin{align}
&f^0_B(w)=m_0^{3/2}\,{\mathrm {sign}}(w)\,e^{-m_0\,|w|}\,,\nonumber\\
&f^0_E(w)=-{m_0^{1/2}}\,e^{-m_0\,|w|}
\end{align}
whereas the massive modes are given by
\begin{align}
&f_B^\mu(w)= m_0\,\sqrt{\frac{\mu}{\pi\,\kappa'}}\,{\rm {sign}}(w)\,\sin\left(\kappa'\,|w|+\alpha\right) \\
&f_E^\mu(w)=\frac{1}{m_0}\,\sqrt{\frac{\mu\,\kappa'}{\pi}}\,\cos\left(\kappa'\,|w|+\alpha\right) 
\end{align}
where, by imposing continuity of $f_E$ and of $f_B$ at $|w|=w_0$ we obtain that $\tan\alpha=-\sqrt{\mu^2/m_0^2-1}$.

\subsubsection{Scalars}

As we seen above there are no massless physical scalars in the spectrum. The wave function of the massive scalars vanishes at $w=w_0\to 0$, as shown by eq.~(\ref{scalarmassive}). Therefore these modes do not couple to matter, and we can ignore their effect.

\subsection{Newton's constant}

As we has seen above, for matter localized on a $\delta$-like brane only the tensors $\gamma_{\mu\nu}$ are sourced. As a consequence we can read the four-dimensional effective four-dimensional Planck mass $M_4$ by computing the coupling between the canonically normalized zero mode of the tensors and the brane stress-energy tensor.  If $\gamma_{\mu\nu}^c$ is the canonically normalized fluctuation of the graviton, then its coupling to matter reads
\begin{equation}
S^{(4D)}_{\rm {int}}=\frac{2}{M_4}\int d^4x\,\gamma_{\mu\nu}^c\,T^{\mu\nu}\,.
\end{equation}

Using the results from section III, we find that the zero mode of $\gamma_{\mu\nu}$ is related by the canonically normalized graviton by $\gamma_{\mu\nu}^c=(M_5^{3/2}/2)\,\gamma_{\mu\nu}^{(0)}$, so that we read the effective Planck mass as $M_4=M_5^{3/2}/f^{(0)}_\gamma(0)=M_5^{3/2}/m_0^{1/2}$.

\section{Conclusions and discussion}

In this paper we  have investigated the possibility of localizing gravity on a 3-brane in an infinite five-dimensional Minkowski background. Our construction relied on endowing the mass of the five-dimensional graviton with a non-trivial profile in the extra dimension. At the linearized level, we found that our model is phenomenologically viable. A fine-tuning of the mass profile parameters is required to support a zero mode of the four-dimensional graviton. By using a square well mass profile as an explicit example, we  have been able to compute the mode functions for the tensor, vector, and scalar modes of the four-dimensional graviton. The main non-trivial features of our model are (1) conserved matter on the brane does not couple to the scalar sector zero mode, (2) there are no ghost modes as long as the four-dimensional theory does not present tachyonic modes and (3) there is a potentially strongly coupled mode, the zero mode of the vector, that is however not sourced as long as matter does not leak into the extra dimension. For a $\delta$-like mass profile for the graviton, we have computed the four-dimensional Planck mass as a function of the parameters of the five-dimensional theory.

Several questions are left open. First, one might ask whether there is some  version of the theory for which the existence of a zero mode is guaranteed by some symmetry. One additional question concerns the fate of the zero vector mode. Does it get a kinetic term at the nonlinear level? Does it lead to a ghostlike excitation? Finally, it would be interesting to study the cosmology of this model. Of course, we expect it to be the standard one as long as the Hubble parameter is much smaller than the mass of the first KK mode of the graviton. However, it would be interesting to know what happens at higher energies. While for the analysis presented in this work only the linearized Lagrangian of massive gravity around five-dimensional Minkowski space was necessary, we will need a full nonlinear Lagrangian of massive gravity such as that of~\cite{deRham:2010kj} (see also~\cite{Nibbelink:2006sz}), or some of its extensions -- such at the quasidilaton model of~\cite{D'Amico:2012zv} -- to study, along the lines of e.g.~\cite{Volkov:2012zb}, the cosmology of the system at the level of the full Einstein equations.

{\bf Acknowledgments.} We thank Shinji Mukohyama, Nemanja Kaloper, Marco Peloso and Oriol Pujolas for useful discussions. L.S. thanks the Institut de Physique Th\'eorique  at the CEA in Saclay and the Laboratoire AstroParticule et Cosmologie at the Universit\'e Paris 7 for hospitality during the completion of this work. This work is partially supported by the NSF grant PHY-120598.

\end{document}